\documentclass[reprint,aps,pra]{revtex4-1}

\usepackage{graphicx}
\usepackage{epsfig}
\usepackage{epstopdf}
\usepackage{subfigure,color}
\usepackage{dcolumn}
\usepackage{bm}
\usepackage{amsmath,amssymb}
\usepackage{xcolor}
\usepackage{graphicx,epsfig,epstopdf}
\usepackage{amsmath}
\usepackage{color}
\usepackage{hyperref}
\usepackage{subfigure}
\usepackage{array}

\usepackage{multirow}
\usepackage[utf8]{inputenc}
\usepackage{accents}
\usepackage[T1]{sansmath} 
\usepackage{bm}
\usepackage{IEEEtrantools}

\def\.{\cdot}

\def\##1{{\bf #1\mit}}
\def\_#1{{\bf #1\mit}}
\def\-#1{{\bf #1\mit}}
\def\=#1{\overline{\overline #1}}

\def\e{\begin{equation}}
\def\f{\end{equation}}

\begin{document}
	\newcommand{\red}[1]{\textcolor{red}{#1}}
	\title{Space-Time Metasurfaces for Perfect Power Combining of Waves}
	
	\author{X.~Wang$^{1}$}
	\author{V.~S.~Asadchy$^{2}$}
	\author{S.~Fan$^{2}$}
	\author{S.~A.~Tretyakov$^1$}
	\affiliation{$^1$Department of Electronics and Nanoengineering, Aalto University, P.O.~Box 15500, FI-00076 Aalto, Finland\\
		$^2$Ginzton Laboratory and Department of Electrical Engineering, Stanford University, Stanford, California 94305,  USA} 
	
	\begin{abstract}
In   passive   linear systems, complete combining of powers carried by  waves from several input channels into a single output channel is forbidden by the energy conservation law.
Here, we demonstrate that  complete  combination of both coherent and incoherent plane waves   can be achieved using metasurfaces with properties varying in space and time. The proposed structure reflects  waves of the same frequency but incident at different angles towards a single direction. The frequencies of the output waves are shifted by the metasurface, ensuring  perfect incoherent power combining. 
The proposed concept of power combining is  general and can be applied for electromagnetic
waves from  the microwave   to  optical domains, as well as for waves of other physical nature.
%
		
	\end{abstract}
	\maketitle
	
\section{Introduction}
Combining electromagnetic waves from multiple input ports to a single output port is highly desired in optical and   microwave engineering. It would overcome   power limits of conventional  sources, enabling high-power laser and long-distance free-space communications.
The most straightforward approach for wave combining is based on beam splitters or power dividers~\cite{pozar2011microwave}. However, these passive components   allow complete combination of coherent beams     only when the beams have specific  phase and amplitude relations among them.  This constraint originates from the passivity of the system~\cite{he2017possibility,hanna2016coherent}. 
In general, input ports of a passive   combiner  are not matched and, when excited by incident waves,  parasitic reflections appear. Only when the combiner is coherently illuminated from all the input ports simultaneously, with specific phases of each input wave, the reflections will interfere destructively, and all the input  power can be directed to the output port. 
Such phase- and amplitude-sensitive operation is analogous to that of recently proposed coherent absorbers and lasers~\cite{baranov2017coherent}. However, in most practical scenarios,   incident coherent waves have arbitrary phases and amplitudes and, therefore, the condition of  destructive interference   of parasitic reflections is not satisfied, and the combining efficiency drops significantly.  


For incoherent incident waves, the passivity constraint does not hold since  the waves do not interfere, and the combined output intensity is  merely a   sum of the input intensities. 
This opportunity has been widely used in high-power lasers and electromagnetic waves. 
It is, for example, certainly possible to combine the power of two waves with orthogonal polarizations,  but in this case, the number of incoming waves is limited to two~\cite{uberna2010coherent}.
One can also combine waves with different frequencies,  which is also known as wavelength beam combining. Using dispersive elements (prisms or gratings), beams from different directions propagating at different frequencies can be collimated at the same output direction \cite{minott1987grating,smits1993power,madasamy2009dual}.
For example, using one \cite{belousov1991quasi,chann2005near} or a pair of grating structures \cite{madasamy2009dual}, the beams from an incoherent laser array are overlapped in one direction through   materials with normal dispersion. Nevertheless, the emission wavelengths from each source must be accurately controlled. 
In addition, this method cannot be used to combine low-terahertz and microwave beams with small frequency differences $\Delta f$ because the required size of gratings would be enormous due to scaling  as $1/\Delta f$~\cite{smits1993power,thumm2002passive}.
Therefore, it is highly desirable to have a compact  system to efficiently combine powers from different ports without restrictions on amplitudes, phases, and frequency differences for the input beams. 

In this paper, we propose a method for perfect power combining    for both coherent and incoherent plane waves using  reflective space-time modulated metasurfaces.  
Spatiotemporal metasurfaces recently  demonstrated  great potential for a plethora of applications, such as Doppler-type frequency shift~\cite{shaltout2015time},   isolation~\cite{wang2020theory,cardin2020surface,hadad2015space}, circulation~\cite{shi2017optical},   nonreciprocal phase shifting~\cite{wang2020theory} and beam splitting~\cite{taravati2019dynamic}.  
We show that such metasurfaces can provide complete power combining in reflection for waves incident from several predefined directions   having identical frequencies but arbitrary phases and amplitudes. The spatial modulation in the metasurface ensures that all the incident waves are reflected into a single output channel. Due to the temporal modulation, incident same-frequency plane waves are reflected as waves with slightly different frequencies, allowing perfect  incoherent power combining in the output channel.
We provide several metasurface designs for complete power combining with weak frequency conversion (using slow temporal modulations).
In all the cases, the combined power at the output channel can be captured by a single receiver with a finite bandwidth. 

\section{results}
\label{sec:examples}

In what follows, we describe  periodic metasurfaces    as   multi-port systems characterized by   scattering matrix~$\bar{\bar{S}}$  \cite{asadchy2017flat}. Propagating Floquet harmonics for a given illumination direction  represent orthogonal channels  and are equivalent to   physical ports of a multi-port system. 

\begin{figure}[tb]
	\centering
	\includegraphics[width=0.65\linewidth]{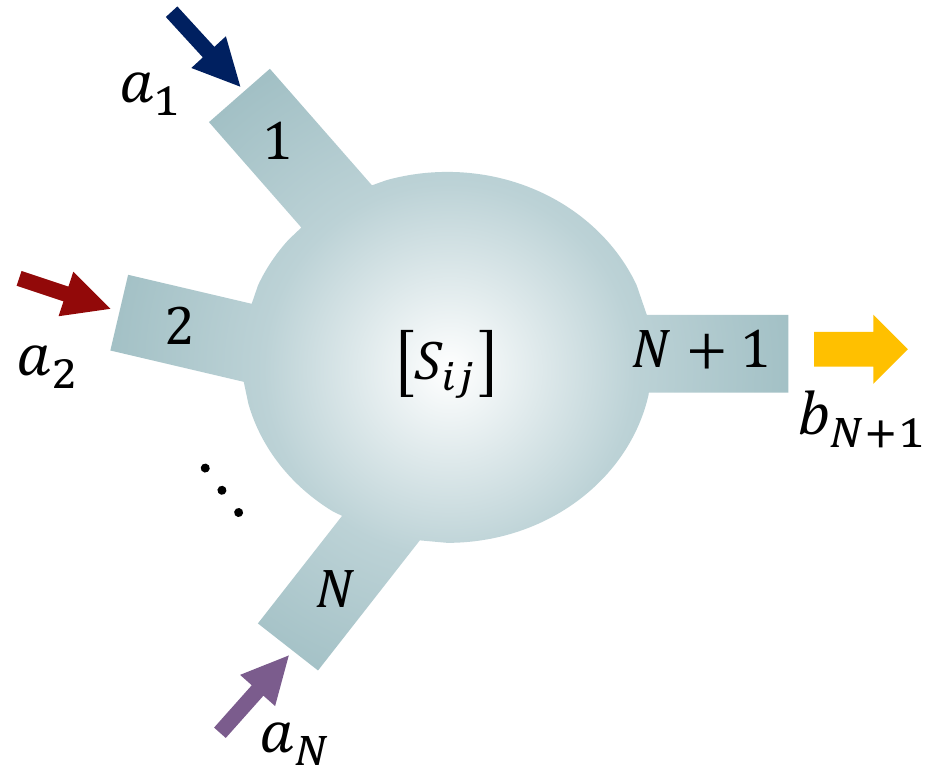}
	\caption{ A general  time-invariant power combiner with $N$ input ports and one output $(N+1)$-st port.  
	} 
	\label{fig: N port}
\end{figure}

First, we show that   linear time-invariant systems (both passive and active) cannot provide complete power combining for coherent plane waves with arbitrary phases and amplitudes. To keep the discussion general, we consider a multi-port system shown in Fig.~\ref{fig: N port}, which can describe power combining in various scenarios: metasurfaces, electrical networks, waveguides,  single scatterers (described by a T-matrix), etc. The multi-port system is excited at input ports~$i$ ($1\leq i\leq N$) with some field amplitudes $a_i=A_i {\rm e}^{j \phi_i}$, where $\phi_i$ are real and  $A_i$ are positive real quantities. The output port~$N+1$ is not excited, that is, $a_{N+1}=0$. From   the scattering  matrix of the system $S$, we can find the output field at this port as $b_{N+1}=  \sum_{i=1}^{N} S_{N+1,i} \, a_i $. The efficiency of power combining expressed   as the ratio of the output power in port~$N+1$ to the total input power at ports from~1 to~$N$ reads
\begin{equation}
    \eta(a_i) = \frac{ \left| \sum_{i=1}^{N} S_{N+1,i} \, A_i \,{\rm e}^{j \phi_i} \right|^2    }{ \sum_{i=1}^{N}   A_i^2 }.   \label{efficiency}
\end{equation}
As   seen from Eq.~(\ref{efficiency}), the combining efficiency strongly depends on the amplitudes   and phases of the input fields, for both passive and active systems.  For any $S$-matrix there in fact exist input fields for which the power combining efficiency is exactly zero. 
For arbitrary input fields $a_i$, the efficiency is always limited by (see  Sec.~1 of the Supplement 1)
\begin{equation}
    \eta(a_i) \leq   \sum_{i=1}^{N} |S_{N+1,i}|^2.   \label{efficiency_limit}
\end{equation}
In active systems, the combining efficiency can be larger than unity for some input fields~$a_i$. On the other hand, in passive systems, the efficiency obeys $\eta(a_i) \leq 1$. The inequality becomes equality only if the following three conditions hold: (i) port~$N+1$ is matched, that is, $S_{N+1,N+1}=0$; (ii) there is no energy dissipation in the system; (iii) all the input fields satisfy the same condition $a_i = C S_{N+1,i}^* $, where $C$ is a complex constant and ``$^*$'' denotes complex conjugate.

In the special case when a passive multi-port network is excited only at one port~$i$  ($1\leq i\leq N$), one can calculate   the average efficiency (averaged over all $N$ ports) of power coupling into the output port~$N+1$ as (see  Sec.~2 of the Supplement 1)
\begin{equation}
    \eta_{\rm coupl}=\frac{1}{N}\sum_{i=1}^{N}  |S_{N+1,i}|^2 \leq \frac{1}{N}.  \label{efficiency_limit2}
\end{equation}

The average efficiency of power combining  can reach its  maximum value   $\eta_{\rm coupl}={1}/{N}$ only in the absence of power dissipation  and if the output port is matched. We see that with increasing the number of input ports~$N$, the average coupling efficiency linearly decreases. Furthermore, complete power combining from multiple input ports becomes exceedingly difficult with increasing~$N$ due to the requirement of mutual coherence of the incident waves.
Obviously, the phase-dependent operation is not desired in practical applications of power combining. In linear time-invariant  systems, the problem cannot be avoided because the monochromatic outgoing waves at the output port always interfere.
\begin{figure}[tb]
	\centering
	\includegraphics[width=1\linewidth]{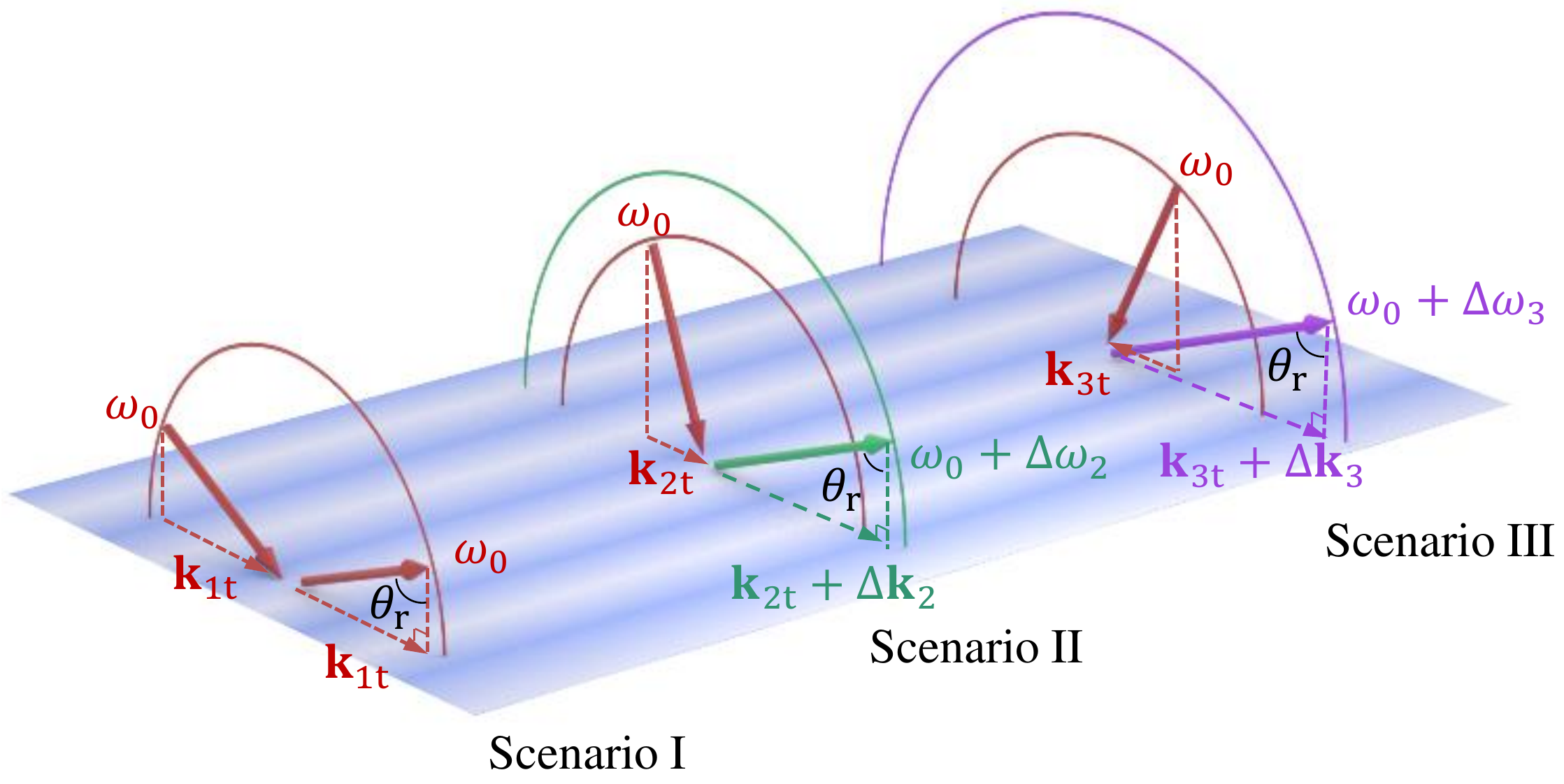}
	\caption{The principle of complete power combining using space-time metasurfaces. The metasurface totally reflects incident plane waves coming  from three different directions (three scenarios) into the same output direction at an angle $\theta_{\rm r}$. The metasurface imparts both tangential spatial and temporal momenta onto incident waves, modifying the isofrequency curves.  
	} 
	\label{fig: principle}
\end{figure}

In order to overcome these fundamental  constraints on power combining, we propose to exploit a metasurface possessing both spatial and temporal modulations of its electromagnetic properties. Such a time-varying multi-port system introduces a frequency shift between the   output waves making them incoherent. Due to this incoherence, the device can fully combine waves \textit{of the same frequency} from multiple input ports to a single output port, and  operate efficiently for both single- and multi-port input waves  with arbitrary amplitudes and phases. The operational principle of the metasurface is  illustrated in   Fig.~\ref{fig: principle}. Plane waves of the same frequency $\omega_0$ incident at different angles on the metasurface (illustrated by three separate scenarios) are completely reflected to the same direction, at an angle $\theta_{\rm r}$. 
The temporal modulation of the metasurface adds different frequencies (temporal momenta) to the three reflected waves: $0$, $\Delta \omega_2$, and $\Delta \omega_3$, respectively. Therefore, the output waves in all the scenarios become incoherent (for coherent inputs) and their powers can be perfectly combined.  The identical reflection angles for the output waves are ensured  by proper spatial modulation of the metasurface: the reflected waves gain different spatial momenta, $0$, $\Delta \mathbf{k}_2$, $\Delta \mathbf{k}_3$, respectively, such that they are deflected towards the same direction.

\begin{figure}[tb]
	\centering
	\includegraphics[width=0.95\linewidth]{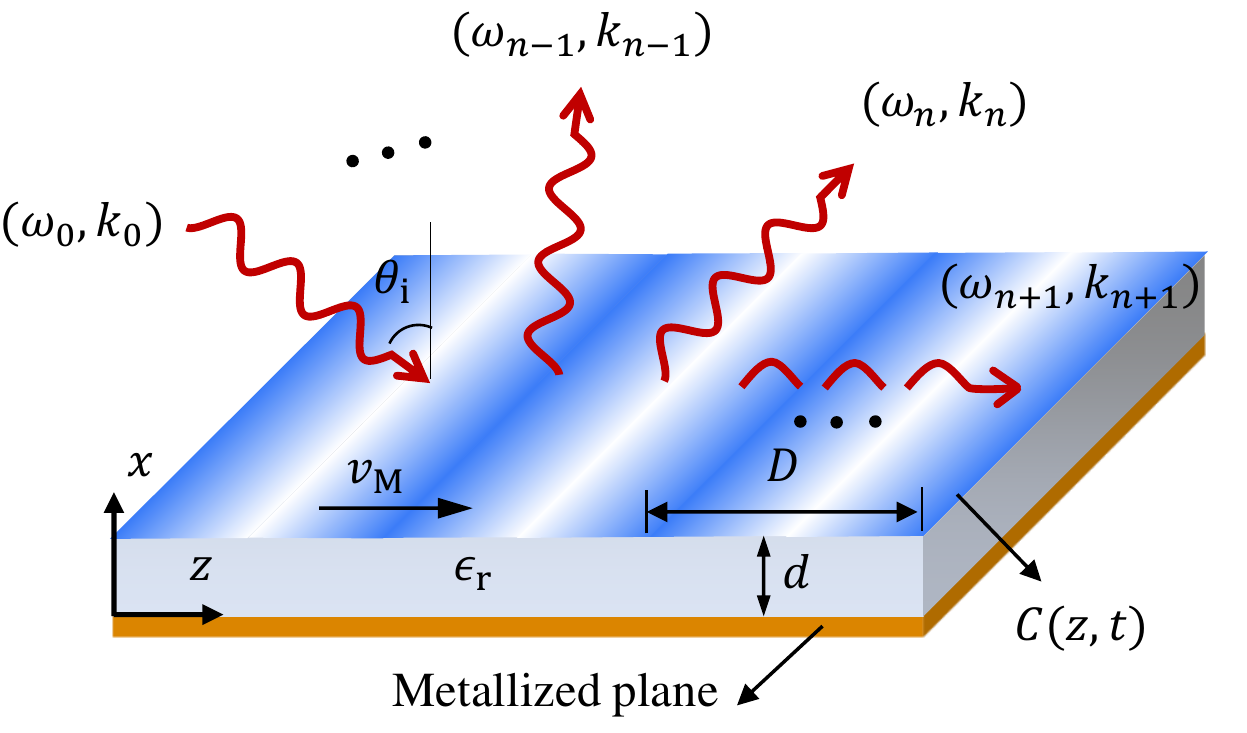}
	\caption{Scattering from a space-time modulated  impedance sheet  positioned on a grounded substrate. 
	} 
	\label{fig:general_scattering}
\end{figure}

We model the metasurface as an impedance sheet positioned on a grounded dielectric substrate, as shown in Fig.~\ref{fig:general_scattering}. Such  geometry was previously exploited both for time-invariant~\cite{wang2018extreme} and time-varying~\cite{wu2020space,li2020time,cardin2020surface} metasurfaces. Here, we model the impedance sheet as a spatiotemporally modulated sheet capacitance $C(z, t)$. Such  an impedance sheet can be realized at microwave frequencies by   a metallic pattern with embedded varactors (voltage-controlled capacitors). For optical frequencies, temporal  modulations of similar type can be realized by voltage-biased $pn$-junction
electrical diodes~\cite{lira2012electrically}, and optically pumping epsilon-near zero materials such as AZO (aluminum-doped zinc oxide)  for ultrafast modulation \cite{kinsey2015epsilon,clerici2017controlling}.
The modulation of surface capacitance has a traveling-wave form with spatial period $D$ (along the $z$-direction)  and temporal  period~$T$, and therefore the modulation velocity is $v_{\rm M}=D/T$. Due to the  periodicity, the modulation function can be expanded into the Fourier series 
\begin{equation}
	C(z,t)=\sum_{m=-\infty}^{+\infty}c_m e^{-jm(\beta_{\rm M} z-\omega_{\rm M}t)},
\end{equation}
where $\beta_{\rm M}=2\pi/D$ and $\omega_{\rm M}=2\pi/T$ are the spatial and temporal modulation frequencies, respectively, and $c_m=c_{-m}^*$ to ensure  $C(z,t)$ is real function.

We consider a TE (transverse electric) plane wave incident from $\theta=\theta_{\rm i}$ at a frequency   $\omega_0$. The incident wavenumber in free space is denoted as $k_0$. Due to the spatial modulation,  the incident wave is scattered into infinitely many spatial harmonics with tangential wavenumbers   $k_{zn}=k_0\sin\theta_{\rm i}+n\beta_{\rm M}$, and due to the time modulation, the propagation frequency of wave in the $n$-th spatial channel is $\omega_n=\omega_0+n\omega_{\rm M}$ \cite{wang2020theory}. The wavevector of the $n$-th harmonic must respect the dispersion relation in free space, $k_n=|\omega_n| \sqrt{\epsilon_0\mu_0}$.
If $|k_{zn}|<|k_n|$, the harmonic is propagating wave and scattered into free space. If $|k_{zn}|>|k_n|$, the harmonic is evanescent.
Once the spatial and temporal modulation  periods, the frequency of the incident waves and the incidence directions are known, the directions of scattering channels (free-space propagating modes) can be determined according to the Floquet theorem. 
When the capacitance function $C(z,t)$ and the substrate properties (permittivity $\varepsilon_{\rm r}$ and thickness~$d$) are chosen, we calculate the amplitudes and phases of all the scattering harmonics   using the mode-matching method (see  Sec.~3 of the Supplement 1).

\begin{figure}[tb]
	\centering
	\includegraphics[width=0.93\linewidth]{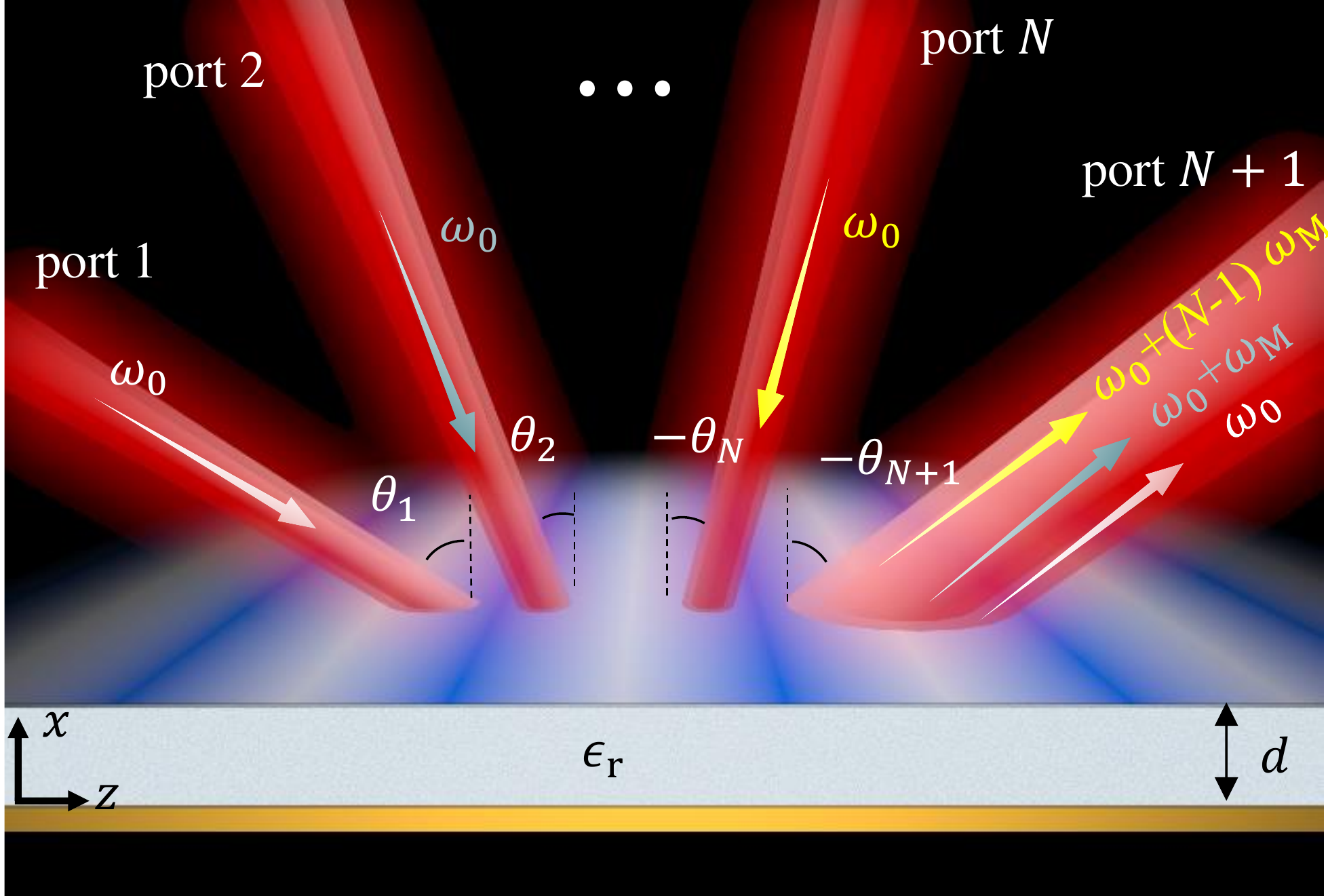}
	\caption{Schematics of a general $N+1$ metasurface power combiner. The output port is along the specular direction of incidence from port~$1$, i.e., $\theta_{\rm 1}=\theta_{N+1}$. All the angles are counted from the metasurface normal   in the anticlockwise direction.
	} 
	\label{fig: N channel}
\end{figure}

The schematics of a general multi-port metasurface combiner is depicted in Fig.~\ref{fig: N channel}, where ports from~$1$ to~$N$ are the input ports, and port $N+1$ is the output port. The directions of the input ports are denoted by the incident angles $\theta_{n}$, where  $n\in[1,N]$ represents the input port index. 
The output port   is along the specular direction of port~1. 
Thus,  when illuminated from port~1, the reflection to port~$N+1$ corresponds to the zero-order harmonic of the metasurface. 
Consecutively, a wave incident from port~$n$ ($n\in[1,N]$)  is reflected to port~$N+1$ as the ($n-1$)-th Floquet harmonic.  According to the Floquet-Bloch theorem, the tangential wavevectors of incidence and its $n-1$ reflection mode are related by expression
\begin{equation}
    k_0\sin \theta_{n}+(n-1)\beta_{\rm M}=\alpha_n k_0\sin \theta_1,  \label{eq: generalized Floquet}
\end{equation}
where $\alpha_n=|\frac{\omega_0+(n-1)\omega_{\rm M}}{\omega_0}|$.
Next, we specify the incidence angles of port~1 and port~$N$ (denoted as $\theta_1$ and $\theta_N$), and thus the spatial period can be determined from Eq.~(\ref{eq: generalized Floquet}):
    \begin{equation}
    D=\frac{(N-1)\lambda_0}{\alpha_N\sin  \theta_1- \sin \theta_N}.\label{eq: generalized D}
\end{equation}
By substituting Eq.~(\ref{eq: generalized D}) into Eq.~(\ref{eq: generalized Floquet}), one can uniquely determine the angles of port~$n$ for $n\in[2, N-1]$:
    \begin{equation}
    \sin \theta_n=
    \alpha_n\sin  \theta_1-\frac{(n-1)(\alpha_N\sin  \theta_1-\sin \theta_N)}{N-1}.
\end{equation}

As soon as the spatial period, modulation frequency, and the directions of all ports are fixed, one can optimize the space-time modulation profile and achieve full reflection from all input ports to the output port~$N+1$. 
According to the  Manley–Rowe relations \cite[Chap.~11.2]{collin2007foundations}, while the energy in time-varying systems is not conserved, the total number of photons entering a lossless system must be equal to that leaving the system. Since our metasurface does not include energy dissipating components, it will conserve the photon flux. 
Due to the periodicity and infinite extent of  the metasurface, the \textit{normal} components of the total incident photon flux  per unit area $P_{{\rm i},x}/(\hbar \omega_{\rm i})$ and  that of the total reflected photon flux  $P_{{\rm r},x}/(\hbar \omega_{\rm r})$ must be equal for any illumination. Here, $P_x$ denotes the normal component of the Poynting vector, and it is proportional to   $E_{\rm i}^2 \cos \theta_{\rm i}$ for incident and to $E_{\rm r}^2 \cos \theta_{\rm r}$ for reflected  plane waves.
To achieve complete power combining using the metasurface, it is sufficient to ensure that, when illuminated from a single port~$n$ ($n\in[1,N]$), the metasurface will reflect all the photons into port~$N+1$. Therefore, we optimize the metasurface so that for single-port excitation the following relation approximately holds
\begin{equation}
    E_{{\rm r}, N+1}=\sqrt{\frac{\omega_{\rm r}}{\omega_{\rm i}}
    \frac{\cos \theta_{n}}{ \cos \theta_{N+1}}}E_{{\rm i},n}.
    \label{eq: amplitude}
\end{equation}
Here, $\omega_{\rm i}=\omega_0$ is the frequency at the input port~$n$ and
$\omega_{\rm r}=\alpha_n \omega_0$ is the frequency at the output port~$N+1$.
For illumination from multiple ports ($n\in[1,N]$), all the photons will be automatically reflected into port~$N+1$ if (\ref{eq: amplitude}) holds, due to the incoherence of the output waves.

To satisfy (\ref{eq: amplitude}) for ports $n\in[1,N]$, we employ   numerical tools to optimize the metasurface   parameters. 
There are in total~$N$ optimization objectives for a metasurface with a given  substrate thickness and substrate permittivity. 
We    use  `fmincon' function in MATLAB   for optimizing the Fourier coefficients of the modulation function $c_m$ of the metasurface. The reflected fields are calculated using the mode-matching method
(see  Sec.~3 of the Supplement 1).

Next, to demonstrate the universality of the proposed concept based on  an example, we design  a 3-port ($N=2$) metasurface power combiner. Without loss of generality, we specify the angles of ports~$1$ and~$2$ as $\theta_1=+45^\circ$ and $\theta_2=0^\circ$, respectively. The output port~$3$ is at the angle $\theta_3=-\theta_1=-45^\circ$ (all the angles are counted from the metasurface normal   in the anticlockwise direction, as shown in Fig.~\ref{fig: N channel}).
The incident frequency is $\omega_0$ and the modulation frequency is assumed much smaller than the incident frequency (arbitrarily chosen as $\omega_{\rm M}=0.044\omega_0$), such that only ports~$1$, $2$, and~$3$ support propagating waves in free space, and other higher-order harmonics excited at the metasurface are evanescent. The advantage of the slow modulation scenario is that such a metasurface can be relatively simply implemented~\cite{lira2012electrically,cardin2020surface,wu2020space,taravati2017nonreciprocal,hadad2016breaking, correas2018magnetic, guo2019nonreciprocal}. Moreover, the converted frequencies at the output port   only slightly differ  from the incident frequency, so that all power  can be efficiently received even by a receiver  with a small bandwidth. 
Based on these parameters, the spatial modulation period    is  determined from Eq.~(\ref{eq: generalized D}) as $D=1.355\lambda_0$.
The substrate thickness is  $d=8.13\times 10^{-4}\lambda_0$, and $\epsilon_{\rm r}=3.55$. 
To find the proper modulation function of the capacitive impedance sheet  that can realize full photon delivery from ports~$1$ and~$2$ to port~$3$, two optimization objectives are set  in numerical optimization according to Eq.~(\ref{eq: amplitude}). 
The first one ensures full specular reflection from port $1$ to port $3$, i.e., $E_{{\rm r}, 3}=E_{\rm i,1}$. The second objective is $E_{{\rm r}, 3}\approx 1.215E_{\rm i,2}$, which ensures full anomalous reflection from port~$2$ to port~$3$. 
Since the number of objectives is small,  we  introduce only three Fourier terms in the modulation function $c_0$, $c_{\pm 1}$, and $c_{\pm 2}$. One of the optimized solution is 
\begin{equation} 
\begin{split}
C(z,t) & = 0.061[1-0.268\cos(\beta_{\rm M}z-\omega_{\rm M}t) \\
 & -0.01\cos(2\beta_{\rm M}z-2\omega_{\rm M}t)]\cdot \omega_0^{-1} ~\mathrm{[F]}.
\end{split}
\end{equation}
Note that this modulation function is independent from the relative phases of waves input to different ports. 
As soon as  the modulation function is obtained, the amplitudes of all the scattering harmonics for each incidence can be calculated following the mode-matching method.  

\begin{figure}[tb]
	\centering
	\includegraphics[width=0.93\linewidth]{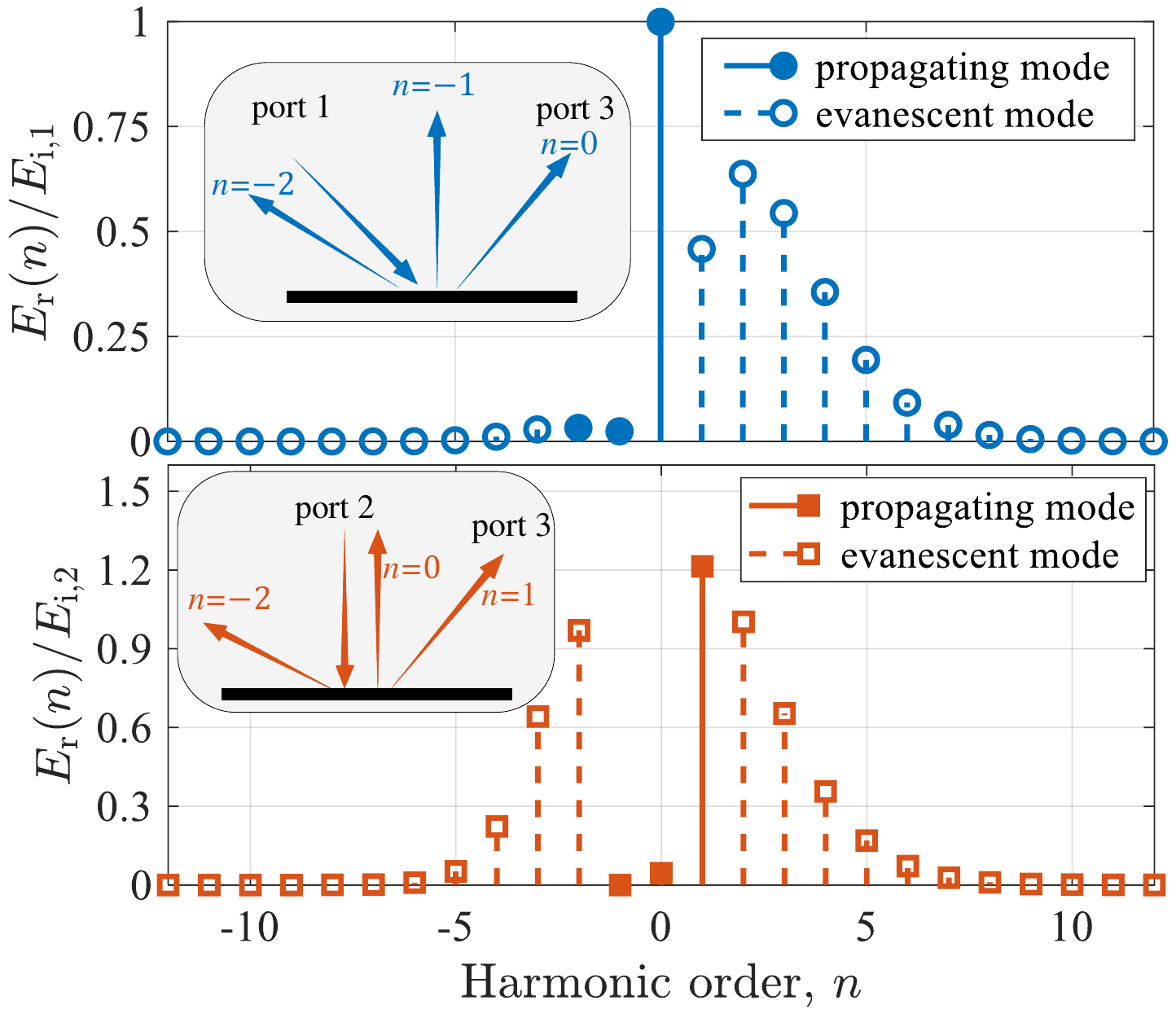}
	\caption{Normalized amplitudes of scattering harmonics for illuminations from port~$1$  (top) and port~$2$ (bottom). Here, index $n$ inside the brackets denotes the scattered harmonic orders that correspond  to different ports in the top and bottom subfigures, as depicted in the insets. 
The scattering harmonics of higher orders are negligibly small.
	} 
	\label{fig: scattering_wm_3port}
\end{figure}

Figure~\ref{fig: scattering_wm_3port} shows the scattering harmonics for two illumination scenarios:  from port~$1$ and from port~$2$. One can see that the two objectives defined in the optimization are reached with  negligible errors, and  undesired waves scattered into input ports have negligible amplitudes.
In addition to the realized power combining of propagating modes, strong evanescent modes (shown by dashed lines) are  excited at the metasurface, but they do not carry power in the normal direction. Nevertheless,  the excitation of evanescent modes is essential  to satisfy the boundary conditions that simultaneously ensure desired   scattering  for both illumination scenarios.
Knowing the amplitudes of all the scattering harmonics, the photon flux reflectance  (the ratio of the normal components of the reflected and incident    photon fluxes per unit area) at each port can be calculated as $R=E_{\rm r}^2\cos \theta_{\rm r}\omega_{\rm i}/(E_{\rm i}^2 \cos\theta_{\rm i}\omega_{\rm r})$. Table~\ref{tab:my-table1} shows the calculated values. 
As is seen, almost all the incident photons are reflected to port~$3$ for illuminations from ports~$1$ and $2$. For completeness, the last column of the table presents the results for the case when the metasurface is illuminated at port~$3$. The total number   of   photons is conserved  for each illumination scenario (the sum of reflectances in each table column equals unity), as expected for a lossless time-modulated metasurface.  Generally speaking, the designed metasurface can be   described by a  normalized scattering matrix whose entries relate the input and output tangential electric fields normalized by the   square roots of the corresponding  frequencies~\cite{wang2021photonic}.
Since the metasurface is lossless, this scattering matrix is unitary. Since the metasurface exhibits both spatial and temporal modulations, the   scattering matrix has an infinite number of elements. The truncated matrix for the designed metasurface can be found in  Sec.~4 of the Supplement 1. 
\begin{table}[tb]
\caption{Photon flux reflectances from the metasurface modulated at $\omega_{\rm M}=0.044\omega_0$. $R_{pq}$ is the reflectance for incidence from port $q$ to port $p$.}
\label{tab:my-table1}
\begin{tabular}{ p{2.4cm}p{2.4cm}p{2.4cm}}
\hline
Illumination from port 1              &  Illumination from port 2         & Illumination from port 3               \\ \hline
$R_{11}=0.087\%$    & $R_{12}=0.000\%$ & $R_{13}=8.365\%$ \\ 
$R_{21}=0.087\%$    & $R_{22}=0.197\%$& $R_{23}=73.508\%$                    \\ 
$R_{31}=99.826\%$   & $R_{32}=99.803\%$ &  $R_{33}=18.127\%$                    \\ \hline
\end{tabular}
\end{table}

\begin{figure}[tb]
	\centering
	\includegraphics[width=1\linewidth]{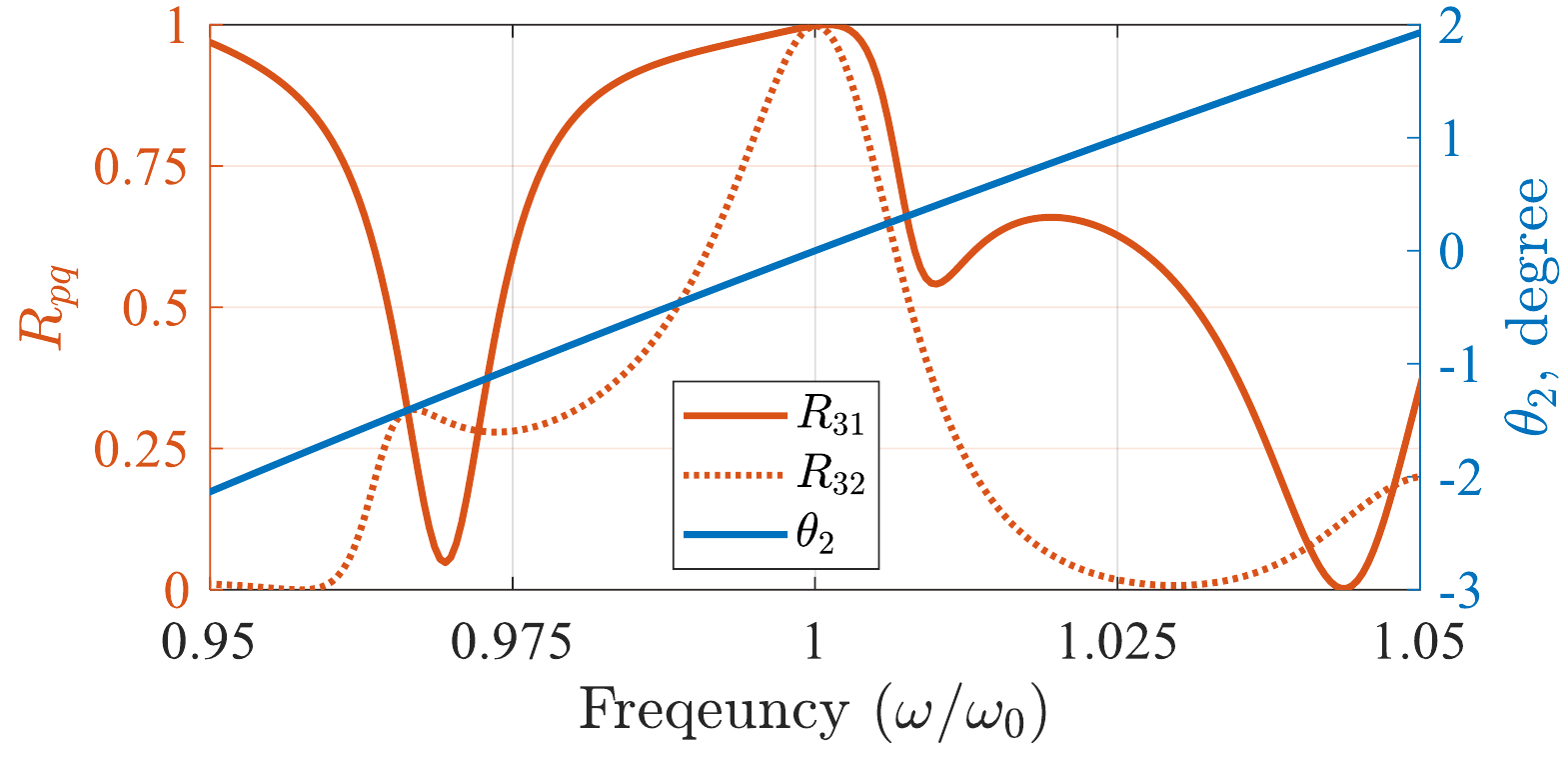}
	\caption{Left axes:   photon flux reflectance from ports~1 and~2 to port~3. Right Axes: variation of the angle of the port~2 direction as the incident frequency changes. 
	} 
	\label{fig: Spara_wm_3port}
\end{figure}

An important characteristic of a power combining device is its bandwidth. 
Figure~\ref{fig: Spara_wm_3port} shows the reflection efficiencies from ports~$1$ and~$2$ to port~$3$, when the incident frequency is swept around $\omega_0$.
It should be noted that, when the incident frequency changes, the direction of port~$2$ also changes according to Eq.~(\ref{eq: generalized D}) because the modulation parameters ($D$ and $\omega_{\rm M}$) and $\theta_{1}$ are assumed to be constant.
From Fig.~\ref{fig: Spara_wm_3port} one can estimate the relative bandwidth of the metasurface to be in the range of 2\%. 
Increasing the modulation speed can widen the bandwidth of the metasurface \cite{williamson2020integrated}.

Modulation of the metasurface  capacitance can be replaced by modulation of  the   permittivity of a material film. At optical frequencies,  permittivity modulation  can be achieved based on the nonlinear Kerr effect using a strong light pump~\cite{guo2019nonreciprocal}. Alternatively, one can discretize the required permittivity in space and apply local voltage control at each sub-cell \cite{lira2012electrically}, which is also used in the microwave range~\cite{hadad2016breaking,taravati2017nonreciprocal}. In the Supplement 1 (Sec.~6), we show that by discretizing one spatial period into at least   ten sub-cells, the above calculated combining efficiency can be achieved.

The above results show power combining from two input ports. The proposed method can be generalized to design power concentrators with an arbitrary number of input ports. As the numbers of ports increases, the numbers of objectives in optimization also increases, and one needs to introduce more Fourier terms in the modulation function. Alternatively, one can introduce additional reactive components in the surface impedance to increase the design freedom. Section~5 of the Supplement 1,  includes a design of a 4-port power combining metasurface.

\section{conclusion}

In summary,  we have introduced the concept of perfect power combining using space-time metasurfaces. Using proper traveling modulation in the metasurface, arbitrary numbers of incident plane waves can be redirected towards one output port with nearly 100\% efficiency.
 The presented method is  general and can be applied from microwave to optical frequencies. Microwave power combining metasurfaces can be realized based on varactors (capacitors with varying capacitance) controlled by an external time-varying voltage. Such implementations of different devices were reported in~\cite{lira2012electrically,cardin2020surface,wu2020space,taravati2017nonreciprocal,hadad2016breaking}. 
In optical frequencies, the traveling-wave modulation can be achieved based on temporally modulating two-dimensional materials such as graphene~\cite{wang2020theory,correas2018magnetic}, epsilon-near-zero materials such as AZO (aluminum-doped zinc oxide),  \cite{kinsey2015epsilon,clerici2017controlling},   as well as nonlinear materials exhibiting Kerr effect \cite{guo2019nonreciprocal}.

Since the designed metasurface power combiner has a finite bandwidth, it is possible to extend the concept to complete and phase-independent combination of incident light pulses, provided that their  spectrum is within the device bandwidth. Perfect pulse combination  is essential for  obtaining high-intensity laser   sources and has important applications for laser-plasma accelerators~\cite{leemans2009laser,esarey2009physics}, particle  sources (e.g., ion)~\cite{haseroth1996physical},  secondary radiation generation~\cite{daido2012review}, and  so on. As is seen from Fig.~\ref{fig: Spara_wm_3port}, the incident angles $\theta_2$ for different frequency components of the pulse  at  port~$2$ must be different to provide high combination efficiency.  Such spatial dispersion of the incident pulse can be generated using conventional techniques with additional diffraction elements~\cite{zhou_coherent_2017}.

	\begin{acknowledgments}
		This work was supported in part by the  Academy of Finland (project 330260)  and the U.S. Air Force Office of  Scientific Research MURI project (Grant No. FA9550-18-1-0379).
	\end{acknowledgments}


\bibliography{references1}

\end{document}